\documentstyle[aps,epsfig]{revtex}
\begin{document}

\title{Seeing phi meson through the dilepton spectra in heavy-ion collisions}

\author{W.S. Chung and C.M. Ko}

\address{Cyclotron Institute and Department of Physics,
Texas A\&M University, \\
College Station, Texas 77843, U.S.A.}

\author{G.Q. Li}
\address{Department of Physics and Astronomy, State University of
New York at Stony Brook,\\
Stony Brook, New York 11794, U.S.A.}

\maketitle

\begin{abstract}

Dilepton spectra from the decay of phi mesons produced in heavy-ion
collisions at SIS/GSI energies ($\sim 2$ GeV/nucleon) are studied
in the relativistic transport model. We include phi mesons produced
from baryon-baryon, pion-baryon, and kaon-antikaon collisions. The
cross sections for the first two processes are obtained from an
one-boson-exchange model, while that for the last process is taken
to be the Breit-Wigner form through the phi meson resonance. For
dileptons with invariant mass near the phi meson peak, we also
include contributions from neutron-proton bremsstrahlung, pion-pion
annihilation, and the decay of rho and omega mesons produced in
baryon-baryon and meson-baryon collisions. Effects due to medium
modifications of the kaon and vector (rho, omega and phi) meson
properties are investigated. We find that the kaon medium effects
lead to a broadening of the dilepton spectrum as a result of the
increase of phi meson decay width. Furthermore, the dropping of phi
meson mass in nuclear medium leads to a shoulder structure in the
dilepton spectrum besides the main peak at the bare phi meson mass.
The experimental measurement of the dilepton spectra from heavy-ion
collisions is expected to provide useful information about the phi
meson properties in dense matter.

\end{abstract}

\newpage

\section{introduction}

The goal of relativistic heavy ion collisions is to study the
properties of nuclear matter under extreme conditions, including
extremely high density and/or temperature. This information is
important for understanding the physics of the early universe and
for studying the dynamic and static properties of stellar objects,
like the structure of neutron stars and supernova explosion. It has
been suggested that there may exist various phase transitions in
the nuclear matter, such as the liquid-gas phase transition
\cite{liqgas}, the kaon condensation \cite{kap86,lee95}, the
restoration of chiral symmetry \cite{brown96}, and the formation of
the quark-gluon plasma \cite{mull96}. These phase transitions, if
they do occur in heavy ion collisions, will have significant
phenomenological implications. For example, as a precursor to
chiral symmetry restoration, hadron properties are expected to be
modified in hot and dense matter \cite{brown91}, leading to
significant effects on their production rate. To study the
properties of hot and dense nuclear matter, heavy-ion accelerators
have been constructed since mid-70s. At the relativistic energy
regime, important facilities include the BEVALAC at Berkeley, the
SIS at Darmstadt, the AGS at Brookhaven and the SPS at CERN. Still
under construction are two ultra-relativistic colliders; the RHIC
at BNL and the heavy ion program at CERN LHC.

To extract information about phase transitions and hadron in-medium
properties from heavy ion experiments, different observables that
are sensitive to the underlying theoretical parameters have been
proposed. Among the important ones are the collective flow of
various types \cite{reis97} and particle production
\cite{cass90,koli96}. Dileptons, because of their relatively weak
interaction with the dense environment, are particularly useful for
studying the medium effects and phase transitions in heavy-ion
collisions \cite{shur78,kkmm86,gale87,liko95}.

Dielectron production in the energy regime of 1-2 GeV/nucleon was
studied by the DLS collaboration at the BEVALAC in Berkeley
\cite{dlsold,dlsnew}. Because of the high hadron multiplicity the
heaviest systems measured with DLS were reactions involving Ca+Ca.
Also, the mass resolution of the DLS spectrometer is not sufficient
to resolve the $\omega$ and $\phi$ peaks from the $\rho$
distribution. A second generation detector HADES is currently under
construction at GSI \cite{hades}. With its high counting rate
capability and large geometrical acceptance, HADES is able to
measure dielectron pairs for the heavy system U+U. With an
invariant mass resolution better than 1\%, a clear identification
of the $\omega$ and $\phi$ peaks can be achieved.

At higher SPS energies, dilepton spectra have been measured by
three collaborations: the CERES collaboration is specialized in the
mass region up to about 1.5 GeV \cite{ceres}, the HELIOS-3
collaboration has measured dimuon spectra from its threshold up to
the $J/\Psi$ region \cite{helios}, and the NA38/NA50 collaboration
measures dimuon spectra in the intermediate- and high-mass region
\cite{na38}. Recent observation of low-mass dilepton enhancement in
heavy-ion collisions by the CERES and HELIOS-3 collaborations has
led to the suggestions of various medium effects
\cite{likob95,cass95,rapp97}, including the dropping of vector
meson masses in hot dense matter.

Theoretical studies on dilepton production at BEVALAC energies were
carried out by Xiong {\it et. al.} \cite{xiong90} and Wolf {\it et.
al.} \cite{wolf90}. They calculated the dilepton yield from
proton-neutron ($pn$) bremsstrahlung, $\Delta$ Dalitz decay, $\pi
\pi$ annihilation, pion annihilation on nucleon, and, in Ref.
\cite{wolf90}, also the $\eta$ Dalitz decay. Their results indicate
that the $\pi^{0}$ Dalitz decay dominates the invariant mass
spectrum below the pion mass. The $\eta$ Dalitz decay then
dominates the spectrum up to an invariant mass of 450 MeV. Above
500 MeV, the most important channels are found to be $pn$
bremsstrahlung and $\pi \pi$ annihilation. These studies reproduce
well the original DLS data \cite{dlsold}. Recently published data
from the DLS collaboration \cite{dlsnew} show, however, a strong
enhancement in the $\eta$ mass region compared with the original
data, implying that the $\eta$ yield could be enhanced. This is,
however, in conflict with the $\eta$ yield and spectra measured by
the TAPS collaboration for similar reactions \cite{taps97}.

In this paper, we will concern ourselves mainly with dilepton
spectra from phi meson decay in heavy-ion collisions at energies
available from the SIS at GSI, where experimental data will become
available in the near future from the HADES collaboration. The
study of phi meson in heavy ion collisions is interesting in view
of the following considerations:

\begin{itemize}
\item{First of all, phi meson is a pure $s\bar s$ state, and its
production in hadronic interactions are therefore suppressed by the
OZI rule \cite{ozi}. This has led to the suggestion of identifying
phi meson enhancement as a possible signature for the formation of
the quark-gluon plasma in heavy-ion collisions \cite{shor85}.}

\item{Phi meson decays chiefly into a kaon-antikaon pair. Since
the mass of phi meson is very close to twice the kaon mass
($m_\phi- 2m_K$ = 0.032 GeV), its width in free space is small
(about 4.4 MeV). In nuclear medium, both kaon and phi meson masses
may change. If the decrease of $m_K^*+m_{\bar K}^*$ is larger than
that of $m_\phi^*$, then the phi meson decay width would increase,
leading to a broader phi meson mass spectrum. On the other hand, if
the decrease of phi meson mass is larger than that of
$m_K^*+m_{\bar K}$, its strong decay channel will be prohibited in
the medium. The phi meson mass spectrum from heavy-ion collisions
thus provides important information on the medium modification of
both kaon and phi mesons. The suggestion of studying the kaon
medium effects from phi meson production in heavy-ion collisions
was first proposed by Shuryak and collaborators \cite{shur91}.}

\item{In QCD sum-rule calculations \cite{hat92}, the change of phi
meson mass is related to the nucleon strangeness content. Thus,
the detection of in-medium phi meson mass is expected to provide
indirect information about the nucleon strangeness content.}
\item{Experimentally, phi meson can be detected from both its $K^+K^-$
and dilepton decay channels. The main difference between these two
channels is that in the $K^+K^-$ case, the strong kaon and antikaon
final-state interactions basically limit the detection of only
those phi mesons that decay at and after freeze-out, while from the
dilepton channel one can also detect phi mesons that decay inside
the initial hot dense matter, i.e., before freeze-out. Therefore, a
simultaneous measurement of the phi meson spectra in the $K^+K^-$
and dilepton channels will provide useful information about the
relative decay probabilities of phi mesons inside and outside the
matter.}
\end{itemize}

Phi meson production from heavy-ion collisions has already been
studied at various energies. At SPS energies it was measured by the
NA38 collaboration \cite{na38fi} and the HELIOS-3 collaboration
\cite{heliosfi} via the dimuon invariant mass spectra. A factor of
2 to 3 enhancement in the double ratio $( \phi / (\omega+\rho^{0}
))_{SU(W)} /( \phi / (\omega+\rho^{0}))_{pW}$ was observed. Various
theoretical attempts have been made to understand this enhancement
\cite{kosa,greiner,koch90}. In particular, an enhancement of the
phi meson yield may be a signature of the formation of a
quark-gluon plasma in the collisions \cite{shor85}. However, the
enhancement can also be explained in hadronic models if one takes
into account the reduced phi meson mass in medium \cite{kosa} or
the formation of color ropes in the initial stage of the collisions
\cite{greiner}.

The phi meson yield has also been measured at AGS/BNL by the E802
collaboration in central collisions between a 14.6 AGeV/c Si beam
and a Au target \cite{e802}. They are identified from the invariant
mass spectrum of $K^{+} K^{-}$ pairs, and the measured phi meson
mass and width are found to be consistent with those in free space.
The ratio of the phi meson yield to the $K^{-}$ yield is about
10\%, which can be understood if thermal and chemical equilibrium
with a temperature of about 110 MeV are assumed at freeze out
\cite{bs95,fireball}. On the other hand, calculations based on the
coalescence model \cite{dover}, in which the phi meson is formed
from the kaon and antikaon at freeze out, underestimate the data by
a large factor, indicating that processes other than kaon-antikaon
annihilation should dominate phi meson production at these
energies.

Phi meson production from heavy-ion collisions at SIS/GSI energies
is being studied by the FOPI collaboration \cite{fopi} through the
$K^{+} K^{-}$ invariant mass distribution. A total of $30\pm8$
$\phi$ has been reconstructed in the reaction Ni+Ni at 1.93
GeV/nucleon from an event sample of $7 \times 10^{6}$ events. Based
on these preliminary results it has been concluded that the phi
meson yield is about 10\% of the $K^{-}$ yield, which is very
similar to that observed at the AGS energies. This is somewhat
surprising since the SIS energies are below the phi meson
production threshold in the nucleon-nucleon collision in free
space, while the AGS energies are well above the threshold. In Ref.
\cite{chung97}, we have studied phi meson production from heavy-ion
collisions at these energies through its $K {\bar K}$ decay
channel. Comparison of the phi meson yield from the transport model
with the preliminary data from the FOPI collaboration \cite{fopi}
seems to indicate that medium effects on phi meson play a
non-negligible role. The present work is a continuation of Ref.
\cite{chung97}. Our main motivation is to examine the feasibility
of detecting directly the reduction of phi meson mass in nuclear
medium from the dilepton invariant mass spectrum from heavy ion
collisions, which will be measured by HADES in the near future
\cite{hades}.

Theoretical studies of the dilepton decay of phi mesons in heavy
ion collisions at SIS energies have been carried out in Ref.
\cite{liko95} by assuming that they are only produced from $K \bar
K$ annihilation. It was found that for dileptons with invariant
mass around $m_\phi$, contributions from $\pi \pi$ annihilation and
phi meson decay are comparable. As shown in Ref. \cite{chung97},
$\phi$ mesons can also be produced from baryon-baryon collisions
and pion-baryon collisions. Because of the larger abundance of
pions and nucleons than kaons, these two channels were found to be
more important than the $K \bar K$ channel. One thus expect that
including these contributions in the transport model would raise
the $\phi$ peak in the dilepton mass spectrum to above the
background from $\pi$ $\pi$ annihilation and also other background
studied in Refs. \cite{xiong90,wolf90}. In addition, we will also
consider the background from direct leptonic decay of rho and omega
mesons produced from baryon-baryon and pion-baryon interactions.

This paper is arranged as follows. In Section 2 we review the
theoretical predictions for the properties of vector mesons and
kaon meson in dense matter. In Section 3 we discuss the elementary
cross sections for vector meson and kaon production in
baryon-baryon and meson-baryon interactions. The results and
discussions of dilepton production from heavy ion collisions are
then presented in Section 4. The papers ends with a brief
conclusion in Section 5.

\section{In-medium properties of mesons}

In this section, we review briefly various theoretical predictions
for the in-medium properties of vector mesons and kaons. Details
can be found in a recent review by two of the authors \cite{kkl}.

\subsection{vector mesons}

Various approaches and models have been used to study theoretically
the vector meson masses in nuclear matter. These include the
scaling anatz of Brown and Rho \cite{brown91}, the QCD sum-rule
approach \cite{hat92,jin95,asa93,weise97,mosel97}, the quark-meson
coupling model \cite{saito97}, and the quantum hadrodynamics (QHD)
\cite{jean94,shi94,song95}. All these studies seem to suggest that
decreasing vector meson masses is a generic consequence of chiral
symmetry restoration at high densities and/or temperature.

Studies on vector meson in-medium masses using the QCD sum rules
were first carried out by Hatsuda and Lee \cite{hat92}. In this
approach, the real part of the current-current correlation function
is expressed in terms of the scalar quark and gluon condensates
after using the operator product expansion(OPE) at short distances.
The imaginary part is, on the other hand, parameterized
phenomenologically. Using the dispersion relation to relate the
real and imaginary parts, vector meson masses are found to satisfy
certain sum rules involving the quark and gluon condensates. To
extend this approach to vector mesons in nuclear medium, one needs
to include not only the density dependence of the condensates but
also non-scalar condensates. With a simple delta-function plus
continuum for the rho meson spectral function, Hatsuda and Lee
obtained the following results for the in-medium vector meson
masses \cite{hat92}
\begin{eqnarray}
{m_{\rho,\omega}^* \over m_{\rho,\omega}} &\approx&
          1-(0.16 \pm 0.06) {\rho \over \rho_0}\\
{m_{\phi}^* \over m_\phi} &\approx&
1-(0.15 \pm 0.05) y {\rho \over \rho_0 }.
\label{Vmedmass}
\end{eqnarray}
The uncertainties in the above expressions are due to uncertainties
in the density dependence of the condensates. At normal nuclear
matter density, $\rho_0$, rho and omega meson masses thus drop by
about 20\%. For phi meson, the in-medium mass depends on the
nucleon strangeness content $y$. Taking $y=0.15$, the phi meson
mass is seen to drop by about 2\% at normal nuclear matter density.

\begin{figure}
\begin{center}
\centerline{\epsfig{file=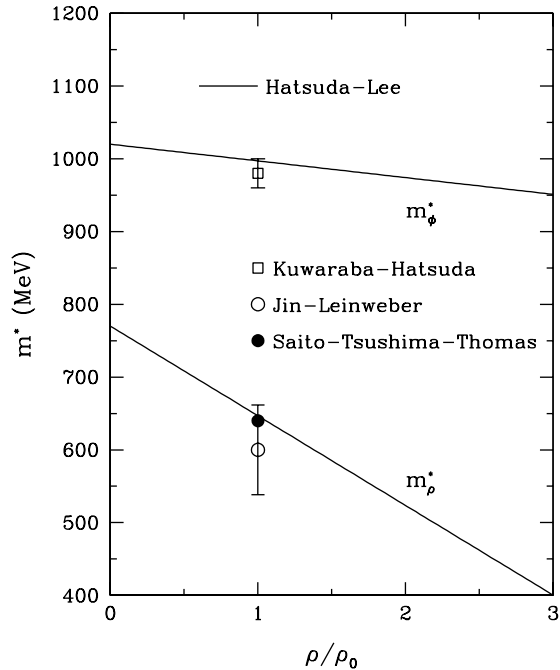,width=4in,height=4in}}
\caption{Vector meson masses as functions of density as obtained in
different theoretical approaches.
\label{vmass}}
\end{center}
\end{figure}

The QCD sum rules for vector mesons were reanalyzed in Ref.
\cite{jin95} by assessing the uncertainties of the condensates and
other inputs using the Monte Carlo error analysis. It was found
that at normal nuclear matter density $m_\rho ^*/m_\rho \approx
0.78 \pm 0.08$, in agreement with the findings of Ref.
\cite{hat92}.

The assumption that the rho meson spectral function in nuclear
medium is the sum of a delta function and a continuum is, however,
too simplistic, as the rho meson has a large decay width in free
space and it also interacts strongly with nucleons. An improved QCD
sum-rule calculation was carried out by Asakawa and Ko \cite{asa93}
using a more realistic rho meson spectral function. They found a
similar decrease of the rho meson mass with density as that of Ref.
\cite{hat92}. More recently, another realistic study of rho meson
spectral function has been carried out in Ref. \cite{weise97}, and
it is found the rho meson mass does not change but its width
becomes larger. This rho spectral function also satisfies the QCD
sum rules. According to Liupold {\it et al.} \cite{mosel97}, the
QCD sum rules do not give a stringent constraint on the rho meson
spectral function as they can be satisfied either with a reduced
mass and width or a larger width but without much change in its
mass.

The quark-meson coupling model has also been used to study the
density dependence of hadron masses \cite{saito97}. In this model,
the vector meson is treated as an MIT bag with two light quarks
which are coupled to the vector and scalar fields generated by the
nuclear medium. It was found that at normal nuclear matter density
the rho and omega meson masses drop by about 17\%, in agreement
with the QCD sum-rule predictions.

Various hadronic models have also been used to study the medium
effect on vector meson masses. Calculations that only include the
polarization of the Fermi sea predict that $\omega$ and $\rho$
meson masses either increase or stay more or less the same in
nuclear matter \cite{chin77,herr93}. The effect of vacuum
polarization or the polarization due to nucleons in the Dirac sea
was studied in Refs. \cite{jean94,shi94,song95} and found to
dominate over the Fermi sea polarization, leading thus to
decreasing rho and omega meson masses. For phi meson, both nucleon
and hyperon vacuum polarization contribute, leading to a reduction
of its mass by about 2-3\% at normal nuclear matter density as in
QCD sum-rule studies \cite{hat95}. In the vector dominance model,
inclusion of dropping kaon-antikoan in-medium mass also leads to a
decrease of phi meson mass in nuclear medium \cite{ko92,klingle97}.

The results from Refs. \cite{hat92,jin95,saito97,hat95} for vector
meson in-medium masses are summarized in Fig. \ref{vmass}. In this
work, as in Refs. \cite{liko95,chung97}, we use the results of
Hatsuda and Lee \cite{hat92} for the density dependence of vector
meson masses and linearly extrapolate it to high densities.

\begin{figure}
\begin{center}
\centerline{\epsfig{file=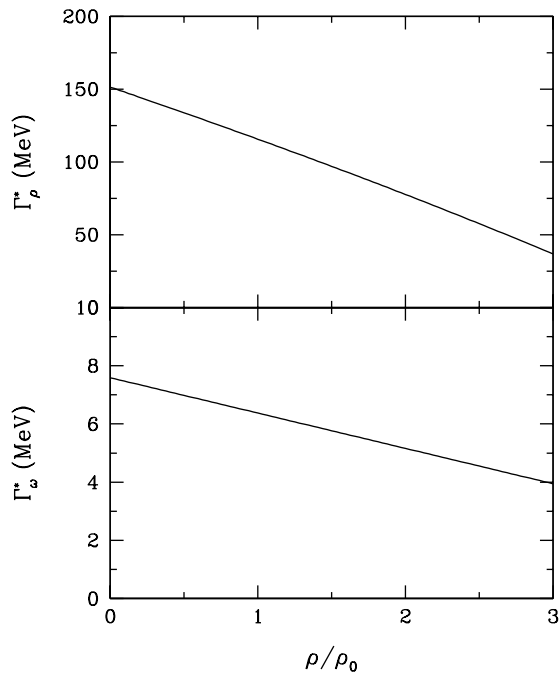,width=4in,height=4in}}
\caption{Rho and omega meson decay widths as a function of
density. \label{vwidth}}
\end{center}
\end{figure}

Because of the change in vector meson masses, their decay widths
are also modified. The rho meson decays dominantly to two pions
with the following decay width in nuclear matter,
\begin{eqnarray}
 \Gamma_{\rho}^{*} = { {g^2_{\rho \pi \pi} } \over {4 \pi} }
     { {1}\over {12 m_{\rho}^{*2}}}
     (m_{\rho}^{*2}-4 m_{\pi}^{2})^{3/2},
\end{eqnarray}
where the coupling $g^{2}_{\rho \pi \pi}/4\pi \approx 2.9$ is
determined from its decay width in free space.

The omega meson decays mostly to three pions. Without a hadronic
model the functional form of the decay width cannot be given. Here
we simply assume that its decay width is proportional to its
in-medium mass, i.e.,
\begin{eqnarray}
\Gamma _\omega ^* \approx \Gamma _\omega {m^*_\omega \over m_\omega}.
\end{eqnarray}
The density dependence of the rho and omega meson decay widths
is shown in Fig. \ref{vwidth}.

\subsection{kaons}

Since phi mesons decay predominantly into $K{\bar K}$ pairs, it is
necessary to address the issue of kaon medium effects. It has been
proposed \cite{shur91} that the phi meson yield can also be used as
a signature for the medium effect on kaons. A reduced kaon
in-medium mass increases $\Gamma_{\phi\rightarrow K{\bar K}}$, so
phi mesons are more likely to decay before freeze out, leading to a
lower yield of final phi mesons that can be detected through their
decay into $K^+K^-$ pairs. A cleaner signature can be observed from
the width of the phi meson peak in the dilepton invariant mass
spectrum. A significantly broadened peak can be another signature
of the kaon medium effect.

Since the pioneering work of Kaplan and Nelson \cite{kap86} on the
possibility of kaon condensation in nuclear matter, a large amount
of theoretical efforts have been devoted to the study of kaon
properties in dense matter, using such diversified approaches as
the chiral Lagrangian
\cite{brown87,wise91,brown94,kai95,lee96,waas97}, the
Nambu$-$Jona-Lasinio model \cite{lutz94}, and the SU(3)
Walecka-type mean-field model \cite{sch94,knor95}. Although
quantitative results from these models are not identical, a
consistent picture has emerged qualitatively; namely, in nuclear
matter the $K^+$ feels a weak repulsive potential, whereas the
$K^-$ feels a strong attractive potential. Most experimental data
for strangeness production and collective flow in heavy ion
collisions at SIS energies \cite{gsiall} have been found to support
the existence of these medium effects
\cite{liall,lilee97,cassall,fae97}. The kaon and antikaon
potentials also have observable effects on kaon azimuthal
distributions \cite{likobrown96} and antikaon flow \cite{liko96}.

\begin{figure}
\begin{center}
\centerline{\epsfig{file=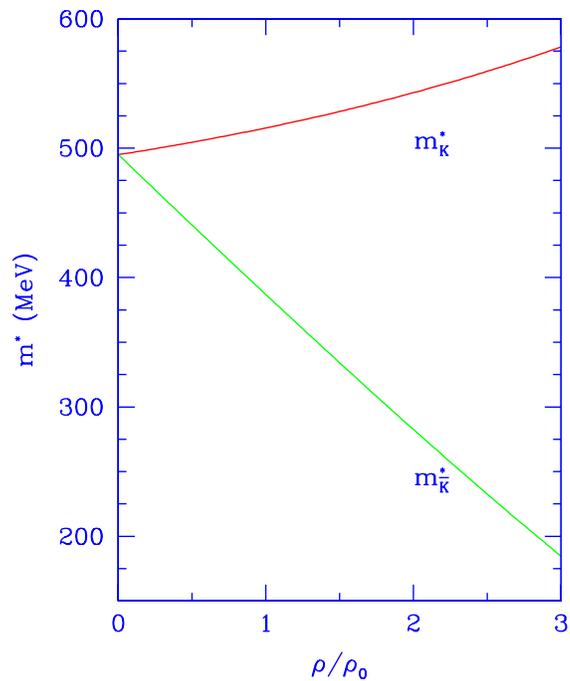,width=4in,height=4in}}
\caption{Kaon and antikaon masses as functions of density
as determined in Ref. \protect\cite{lilee97}.
\label{kmass}}
\end{center}
\end{figure}

We will consider two scenarios for kaon properties in nuclear
medium, one with and one without medium modification. From the
chiral Lagrangian the kaon and antikaon in-medium masses can be
written as \cite{lilee97}
\begin{eqnarray}
m_K^*=\left[m_K^2-a_K\rho_S +(b_K \rho )^2\right]^{1/2} + b_K \rho,
\end{eqnarray}
\begin{eqnarray}
m_{\bar K}^*=\left[m_K^2-a_{\bar K}\rho_S +(b_K \rho
)^2\right]^{1/2} - b_K \rho,
\end{eqnarray}
where $b_K=3/(8f_\pi^2)\approx 0.333$ GeV$\cdot$fm$^3$, $a_K$ and
$a_{\bar K}$ are two parameters that determine the strength of the
attractive scalar potential for kaon and antikaon, respectively. If
one considers only the Kaplan-Nelson term, then $a_K=a_{\bar
K}=\Sigma _{KN}/f_\pi ^2$. In the same order, there is also a range
term which acts differently on kaon and antikaon, and leads to
different scalar attractions. Since the exact value of $\Sigma
_{KN}$ and the size of the higher-order corrections are still under
debate, we take the point of view that $a_{K,{\bar K}}$ can be
treated as free parameters and try to constrain them from the
experimental observables in heavy-ion collisions. In Ref.
\cite{lilee97} it was found that $a_K\approx 0.22$
GeV$^2$$\cdot$fm$^3$ and $a_{\bar K}\approx 0.45$ GeV$^2$fm$^3$
provide a good description of the kaon and antikaon spectra from
Ni+Ni collisions at 1 and 1.8 AGeV. These values will be used in
this work as well. The density dependence of the kaon and antikaon
masses is shown in Fig. \ref{kmass}.

\begin{figure}
\begin{center}
\centerline{\epsfig{file=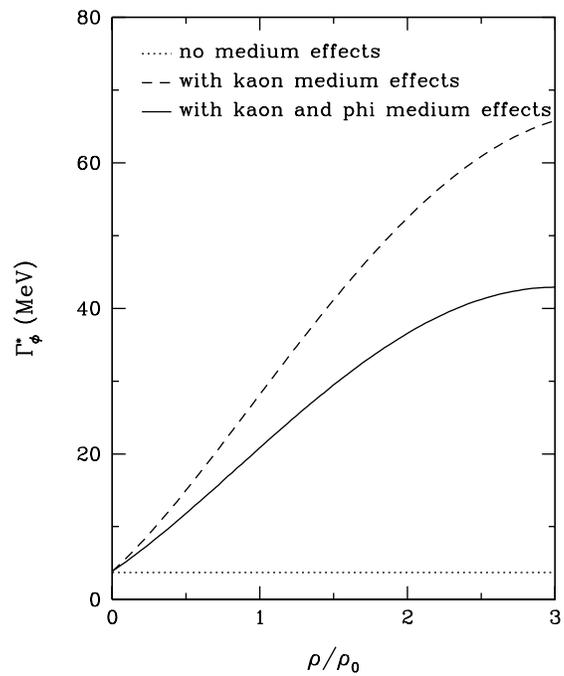,width=4in,height=4in}}
\caption{Phi meson decay widths in three scenarios:
without medium effects (dotted curve), with kaon medium effects
(dashed curve), and with both kaon and phi medium effects (solid
curve).
\label{fiwidth}}
\end{center}
\end{figure}

Because of the change of phi and/or kaon masses with nuclear
density, the phi meson decay width is also modified. According to
Ref. \cite{liko95}, the phi meson decay width is given by
\begin{eqnarray}
\Gamma _\phi ^* = {g_{\phi K {\bar K}} \over 4\pi} {1\over 6 m_\phi
  ^{*5}} \left[(m_\phi^{*2}-(m^*_K+m^*_{\bar K})^2)(m_\phi^{*2}
-(m^*_K-m_{\bar K}^*)^2)\right]^{3/2}.
\end{eqnarray}
In Fig. \ref{fiwidth}, we show the density dependence of the phi
meson decay width in three different scenarios for phi and kaon
masses. Generally, the phi meson decay width is seen to increase with
density if kaon medium effects or both kaon and phi medium effects
are included. In the latter case, the dropping of $m_K^*+m_{\bar
K}^*$ is apparently more significant than that of $m_\phi^*$.

\section{Particle production cross sections}

One of the most important inputs in the transport model for
particle production from heavy-ion collisions is the elementary
particle production cross section in hadron-hadron interactions. In
this section we shall discuss the vector meson and kaon production
cross sections that are used in the present work. We will also
discuss dilepton production cross sections from $pn$ bremsstrahlung
and pion-pion annihilation, as well as the decay widths of vector
mesons into dileptons.

\subsection{phi meson production}

In heavy-ion collisions at SIS energies, the most abundant
particles are nucleons, $\Delta$ resonances, and pions. Phi meson
can thus be produced from reactions such as $NN\rightarrow NN\phi$,
$N\Delta\rightarrow NN\phi$, $\Delta\Delta\rightarrow NN \phi$,
$\pi N \rightarrow \phi N$, and $\pi \Delta \rightarrow \phi N$. In
addition, phi meson can also be formed from kaon-antikaon
annihilation, which has a cross section of usual Breit-Wigner form
\begin{eqnarray}
\sigma (K{\bar K}\rightarrow \phi )
= {3\pi \over k^2} {(m_\phi \Gamma _\phi )^2 \over
(M^2-m_\phi^2)^2 + (m_\phi \Gamma _\phi )^2},
\end{eqnarray}
where $M$ is the invariant mass of the kaon-antikaon pair and $k$
is the magnitude of the kaon momentum in the center-of-mass frame.

The cross sections for the other processes listed above have been
studied in Ref. \cite{chung97} based on an one-boson exchange
model. For the reaction $\pi B \rightarrow\phi N$, where $B$
denotes either a nucleon or a $\Delta$ resonance, rho meson
exchange is used, while both rho and pion exchange are introduced
for the reaction $BB\rightarrow NN\phi$. Coupling constants needed
for evaluating these cross sections are either taken from the Bonn
model for $NN$ potential \cite{mach89} or determined from the
measured width for $\phi\rightarrow\pi\rho$. Most cut-off
parameters at the interaction vertices aer also taken from the Bonn
model. Since the exchanged rho and pion are virtual, two cut-off
parameters at the $\phi\pi\rho$ vertex are introduced, and they are
determined by fitting to available experimental data for the
reactions $\pi^-p\rightarrow \phi n$ and $pp\rightarrow pp\phi$
\cite{xdata}. This model is then extended, without introducing
further adjustable parameters, to calculate the cross sections
involving baryon resonances.

\subsection{rho and omega meson production}

An important background for the dilepton spectrum around phi meson
mass comes from the tail part of the dilepton spectra from the
decay of omega and rho mesons, which are produced from
baryon-baryon and pion-baryon interactions. The rho meson can also
be formed in pion-pion annihilation. In this work, we treated the
dilepton spectrum from pion-pion annihilation using the so-called
form factor approaches \cite{liko95},i.e., without considering the
explicit formation and propagation of rho meson from pion-pion
annihilation.

\begin{figure}
\begin{center}
\centerline{\epsfig{file=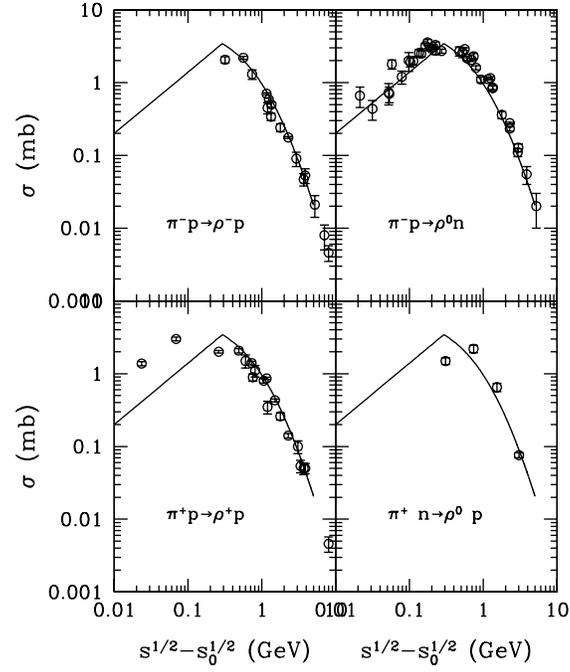,width=4in,height=4in}}
\caption{Comparison of the theoretical parameterization, Eq.
(\ref{para-pirho}), with the experimental data for $\pi N \rightarrow
\rho N$.
\label{pirho}}
\end{center}
\end{figure}

Several parameterizations have been proposed for the rho and omega
meson production cross sections in pion-nucleon interactions
\cite{cug90,sib96,sib97}. These parameterizations have been used in
studying vector meson production in pion-nucleus \cite{weid97} and
proton-nucleus \cite{sib98} reactions through their dilepton
decays. Here we shall follow the one introduced in Ref.
\cite{sib96}. For rho meson production, experimental data for the
four channels, $\pi^+p\to\rho^+p$, $\pi^+n\to\rho^0p$,
$\pi^-p\to\rho^-p$, and $\pi^-p\to\rho^0n$ \cite{xdata}, show that
their cross sections are similar and can be parameterized by
\begin{eqnarray}\label{para-pirho}
\sigma (\pi^- p \rightarrow \rho^0 n)  =&  9.75
(\sqrt s-\sqrt {s_0})^{0.844} ~{\rm mb}, &~\sqrt s\le 2 ~{\rm GeV}
\nonumber\\
=&64.1 s^{-2.11} ~{\rm mb}, &~\sqrt s> 2 ~{\rm GeV},
\end{eqnarray}
where $\sqrt {s_0} = m_N+ m_\rho$ is the threshold. This
parameterization is slightly different from that proposed in Ref.
\cite{sib96}, in order to better fit the data for $\pi ^-
p\rightarrow \rho^0 n$ near the threshold. Comparison of this
parameterization with the experimental data is shown in Fig.
\ref{pirho}.

\begin{figure}
\begin{center}
\centerline{\epsfig{file=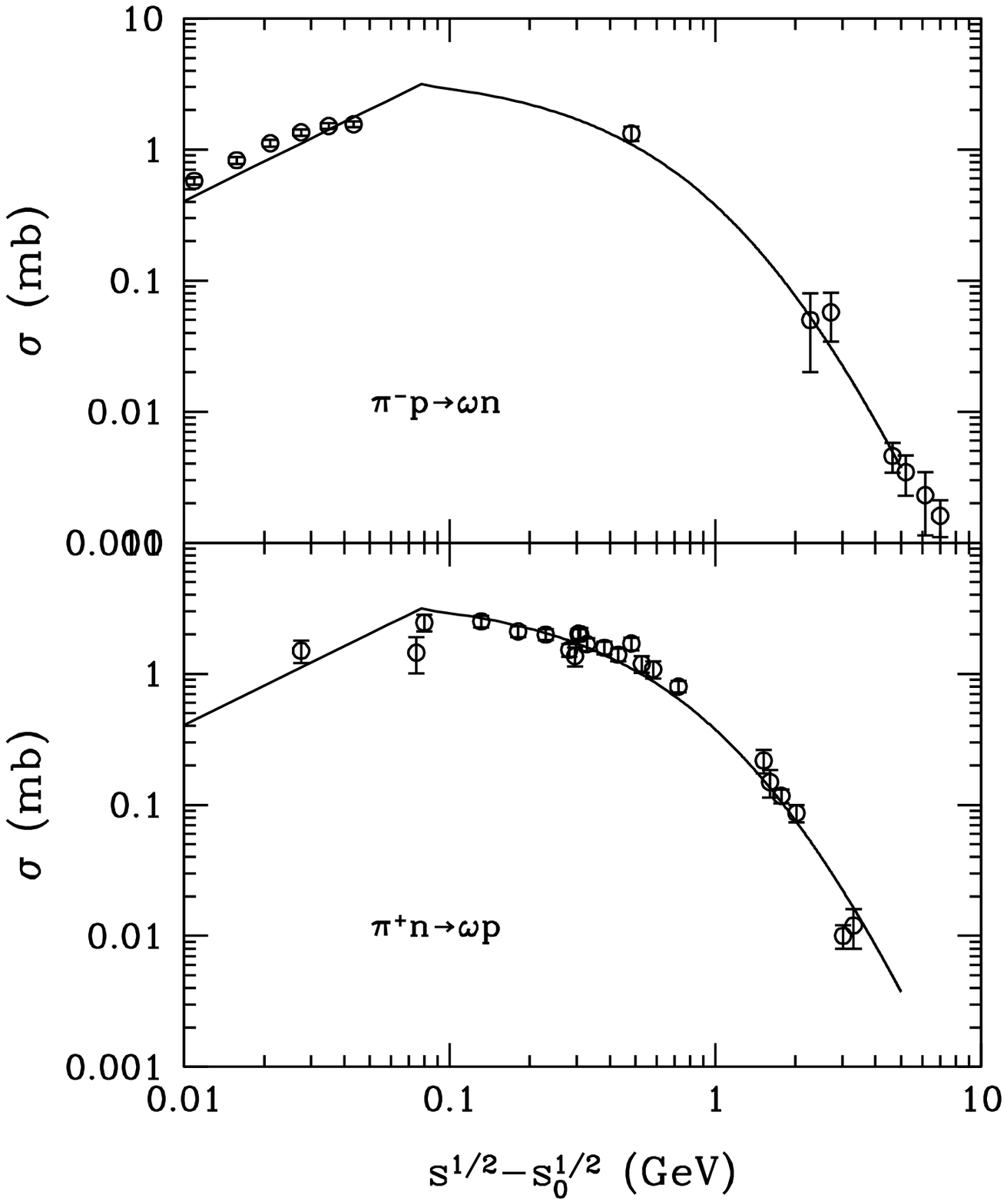,width=4in,height=4in}}
\caption{Comparison of the theoretical parameterization, Eq.
(\ref{para-omega}), with the experimental data for $\pi N \rightarrow
\omega N$.
\label{piom}}
\end{center}
\end{figure}

Similarly, available experimental data for omega meson production
in pion-nucleon collisions can be fitted by the following
parameterization, which is taken from Ref. \cite{sib96},
\begin{eqnarray}\label{para-omega}
\sigma (\pi ^- p \rightarrow \omega n) =& 40.43
(\sqrt s-\sqrt {s_0}) ~{\rm mb}, ~ & \sqrt s\le 1.8 ~{\rm GeV}
\nonumber\\
=& 61.57  s^{-2.55} ~{\rm mb}, ~ &\sqrt s> 1.8 ~{\rm GeV},
\end{eqnarray}
where $\sqrt {s_0} = m_N +m _\omega$ is the threshold. Comparison
of this parameterization with the experimental data is shown in
Fig. \ref{piom}.

\begin{figure}
\begin{center}
\centerline{\epsfig{file=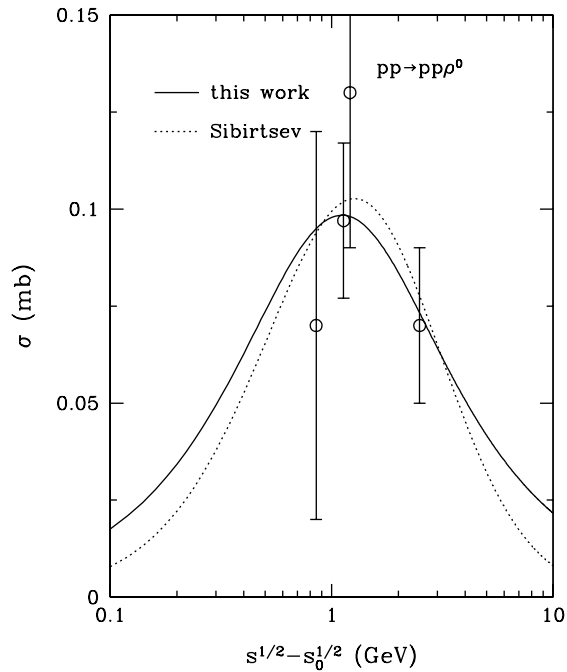,width=4in,height=4in}}
\caption{Comparison of the theoretical parameterizations with the
experimental data for $pp\to pp\rho$. The solid line is from this
work, and the dotted line is from Ref. \protect\cite{sib96}.
\label{pprho}}
\end{center}
\end{figure}

\begin{figure}
\begin{center}
\centerline{\epsfig{file=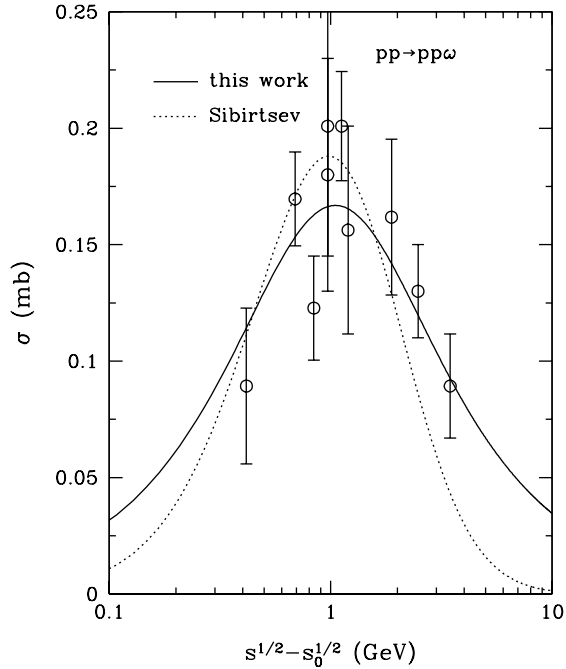,width=4in,height=4in}}
\caption{Same as Fig. \protect\ref{pprho}
for $pp \rightarrow pp\omega$.
\label{ppom}}
\end{center}
\end{figure}

Rho and omega production from the nucleon-nucleon interaction has
been studied in Ref. \cite{sib96} using an one-boson-exchange
model, assuming that these processes can be factorized into a two
step process. A form factor is introduced to account for the
virtuality of the pion. Furthermore, simple parameterizations were
proposed. Here we use a somewhat different form of
parameterization, which has been used also for the parameterization
of the eta meson production cross section in proton-proton
collision near the threshold \cite{wolf90}, i.e.,
\begin{eqnarray}
\sigma ( pp\rightarrow pp\rho^0) = {0.393 (\sqrt s -\sqrt {s_0})
\over 1.05 + (\sqrt s-\sqrt {s_0})^2} ~{\rm mb},
\end{eqnarray}
where $\sqrt {s_0}= 2m_N+m_\rho$ is the threshold,
and
\begin{eqnarray}
\sigma(pp \rightarrow pp \omega ) = {0.219 (\sqrt s- \sqrt {s_0}
\over 1.238 + (\sqrt s - \sqrt {s_0})^2} ~{\rm mb},
\end{eqnarray}
with the threshold $\sqrt {s_0}= 2m_N+m_\omega$. Comparison of
these parameterizations with experimental data and the
parameterizations from Ref. \cite{sib96} are shown in Figs.
\ref{pprho} and \ref{ppom}.

The cross sections for rho and omega production from other isospin
channels of nucleon-nucleon interactions are not available in Ref.
\cite{sib96}. Here we simply assume that are the same as those for
proton-proton interactions. Vector mesons can also be produced in
these processes with nucleons replaced by $\Delta$'s. These cross
sections cannot be studied experimentally. Here we follow the usual
procedure used in transport models by assuming that these cross
sections are the same as those for the corresponding nucleon
channels.

\subsection{kaon and antikaon production}

In heavy-ion collisions at incident energies considered in this
work, kaons can be produced from pion-baryon and baryon-baryon
collisions. For the former we use the cross sections obtained in
the resonance model by Tsushima {\it et al.} \cite{tsu94}. For the
latter the cross sections obtained in the one-boson-exchange model
of Ref. \cite{liko98} are used. Both models describe the available
experimental data very well. For antikaon production from
pion-baryon collisions we use the parameterization proposed by
Sibirtsev {\it et al.} \cite{sib97b}. For baryon-baryon collisions,
we use a somewhat different parameterization, which describes the
experimental data better than Ref. \cite{sib97b}. In addition, the
antikaon can also be produced from strangeness-exchange processes
such as $\pi Y\rightarrow {\bar K}N$, where $Y$ is either a
$\Lambda$ or $\Sigma$ hyperon \cite{ko83}. Cross sections for these
processes are obtained from the inverse ones, ${\bar K}N\rightarrow
\pi Y$, by the detailed-balance relation. All parameterizations for
the elementary cross sections and comparisons with experimental
data can be found in Ref. \cite{lilee97}.

\subsection{final-state interactions}

Particles produced in elementary hadron-hadron interactions in
heavy-ion collisions cannot escape the hadronic matter freely.
Instead, they are subjected to strong final-state interactions. For
kaons, because of strangeness conservation, their scattering with
nucleons at low energies is dominated by elastic and pion
production processes, which do not affect its final yield but
change its momentum spectrum. The final-state interaction for the
antikaon is much stronger. Antikaons can be destroyed in the
strangeness-exchange processes, and they also undergo elastic
scattering. Both the elastic and absorption cross sections increase
rapidly with decreasing antikaon momenta.

Final-state interactions for phi meson were analyzed in Ref.
\cite{chung97}. These include both absorption and elastic
rescattering by nucleons. Of all absorption processes considered,
the $\phi N\rightarrow \Lambda K$, which is not OZI suppressed, was
found to be the most important. In this work we follow Ref.
\cite{chung97} for the treatment of phi meson final-state
interactions.

We also consider the absorption and rescattering of rho and omega
mesons. Reactions such as $\omega N \rightarrow\pi N, ~\pi\Delta $
and $\rho N \rightarrow \pi N, ~\pi\Delta$ are included. Cross
sections for these reactions can be found from those of the
reversed processes and the detailed balance relations. The cross
sections for $\omega N$ and $\rho N$ elastic scattering have been
extracted from the cross sections of the corresponding
photoproduction processes using the photo-dissociation model
\cite{joo67}. It is found to be 5.6 mb at $p_{\rm lab}$ around 3
GeV. As elastic scattering does not change the dilepton yield, its
effect on the dilepton spectrum is not expected to be
significantly. We thus assume that their cross sections take a
constant value in the energy range relevant for our calculations.

\subsection{dilepton production}

A vector meson can decay directly into a lepton pair. We use the
following mass-dependent leptonic decay width for vector mesons
\cite{likob95},
\begin{eqnarray}
\Gamma _{V\rightarrow e^+e^-} (M) = C _{e^+e^-} {m_V^4\over M^3},
\end{eqnarray}
where $M$ is the mass of the vector meson, while $m_V$ is its mass
at the centroid. From the measured widths, the coefficient
$C_{e^+e^-}$ is determined to be $8.814 \times 10^{-6}$,
$0.767\times 10^{-6}$, and $1.334 \times 10^{-6}$ for $\rho^0$,
$\omega$, and $\phi$ decay, respectively. To obtain the final
dilepton spectra from vector meson decays, we need to integrate
over time and add also the contribution after the freeze-out.
Denoting $dP_\phi (t)/dt$ the differential probability of finding
phi mesons at time $t$, the dilepton invariant mass spectrum from
phi meson decays is then given by
\begin{eqnarray}
{dP_{e^+e^-}\over dM} = \int ^{t_f}_0 {dP_\phi (t)\over dM}
\Gamma _{\phi \rightarrow e^+e^-} (M) dt
+{dP_\phi (t_f)\over dM} {\Gamma _{\phi \rightarrow e^+e^-} (M)
\over \Gamma _\phi (M)}.
\end{eqnarray}
The freeze-out time $t_f$ follows automatically in the transport
model when particle collisions become negligible. The first term in
the above expression gives the contribution from phi meson decay
`inside' the fire ball, and the second term gives that from decay
`outside' the fire ball. Similar procedures are used for dileptons
from rho and omega meson decays.

In addition, we shall also include dileptons from $pn$
bremsstrahlung and pion-pion annihilation. Our procedure for $pn$
bremsstrahlung is the same as in Xiong {\it et. al.}
\cite{xiong90}. In the soft-photon approximation and including the
phase-space correction, the differential cross section for
dileptons with invariant mass $M$ is given by:
\begin{eqnarray}
  {d \sigma \over d^3{\bf p}dM} \simeq
    { {\alpha^{2}} \over {6 \pi^{3}} }
    { {\bar \sigma(s)} \over {ME^{3}} }
    { {R_{2}(\surd s_{2})} \over {R_{2}(\surd s)} }
\end{eqnarray}
where
$$
  R_{2}(s)= \sqrt{1-{{4m^2}\over {s}} },
$$
$$
  s_{2}=s+M^{2}-2 E \surd s,
$$
$$
  \bar {\sigma}(s) = { { s-4m^2 } \over {2 m^{2}} }\sigma(s).
$$
In the above, $\alpha$ is the fine structure constant, $E$ and {\bf
p} are the total energy and momentum of the dilepton pair,
respectively, $\sigma(s)$ is the cross section of nucleon-nucleon
elastic scattering, while $m$ is the nucleon mass.

The dilepton production cross section from pion-pion annihilation is
well-known \cite{gale87,liko95}, i.e.,
\begin{eqnarray}
\sigma (M) = {4\pi\alpha^2 \over 3M^2}
\sqrt{1-{ {4 m_{\pi}^2} \over {M^2}} }
\vert F_{\pi} \vert^{2},
\end{eqnarray}
where the pion electromagnetic form factor is given by:
\begin{eqnarray}
\vert F_{\pi} \vert^{2}  =
{ {m_{\rho}^4} \over { (M^{2}-m_{\rho}^2)^{2}
+m_{\rho}^{2} \Gamma_{\rho}^{2} } },
\end{eqnarray}
in terms of the rho meson mass $m_{\rho}$ and width $\Gamma_{\rho}$.

\section{results and discussions}

As in Ref. \cite{chung97}, the dynamical evolution of heavy-ion
collisions at SIS energies is described by the relativistic
transport model, originally developed in Ref. \cite{ko87} and
extensively used in studying medium effects in heavy ion collisins
\cite{koli96}. As we are mainly interested in the possibility of
seeing phi and kaon medium effects on the dilepton spectra, we
shall present our results in the mass region from 0.8 to 1.2 GeV in
central Ni+Ni collisions at 1.93 AGeV.

\begin{figure}
\begin{center}
\centerline{\epsfig{file=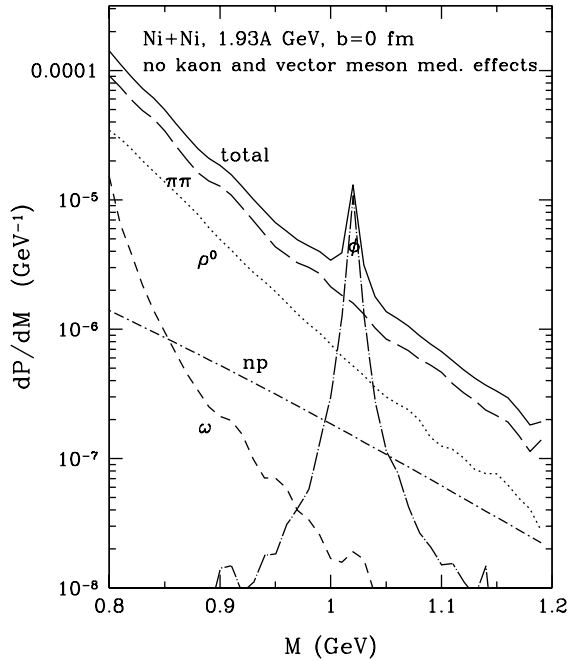,width=4in,height=4in}}
\caption{Dilepton invariant mass spectra in central Ni+Ni collisions
at 1.93A GeV. The results are for the scenario neglecting both the
kaon and vector meson medium effects.
\label{ee}}
\end{center}
\end{figure}

In Fig. \ref{ee} we show the dilepton invariant mass spectra in the
case that both vector and kaon medium effects are neglected. Here,
as well as in the following figures, the histogram bin size is
taken to be 10 MeV, corresponding roughly to the mass resolution of
HADES in the phi meson mass region. The major background around phi
meson peak comes from $\pi \pi$ annihilation. The contribution from
direct rho meson decay is about a factor of 3 below that from
$\pi\pi$ annihilation. Furthermore, the contribution from $pn$
bremsstrahlung is about one order of magnitude smaller than other
background in this mass region. Finally, the contribution from
direct omega meson decay is insignificant due to its narrow mass
distribution. A well defined phi meson peak is seen in the dilepton
spectrum, reflecting its small decay width of about 4 MeV. Overall
the phi meson peak is about a factor 4-5 above the background. This
make its detection relatively easy if the mass resolution is about
1\%.

\begin{figure}
\begin{center}
\centerline{\epsfig{file=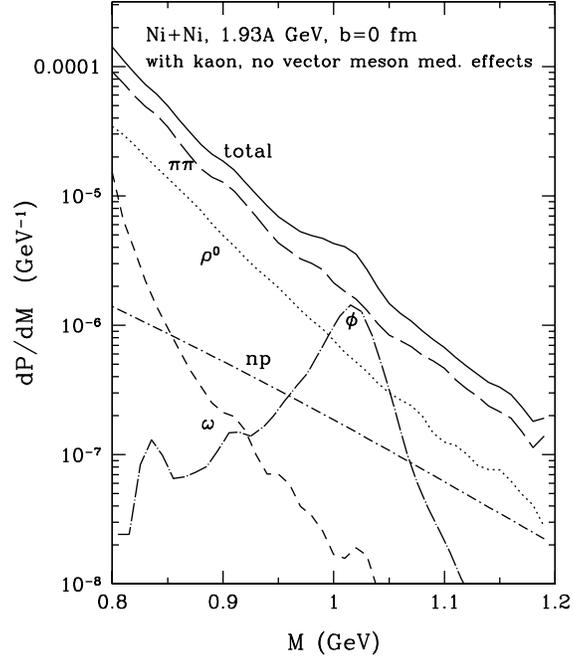,width=4in,height=4in}}
\caption{Same as Fig. \protect\ref{ee}
for the scenario including kaon medium effects but neglecting
vector meson medium effects.
\label{eem}}
\end{center}
\end{figure}

In the second scenario, we turn on the kaon medium effects as were
required to explain the measured kaon yields and flow
\cite{gsiall,liall,lilee97,cassall,fae97}. As shown in Fig.
\ref{fiwidth}, because of the opening-up of the phase space, the
phi meson decay width increases substantially in nuclear medium. As
a result, phi mesons tend to decay faster and therefore less number
of phi mesons can be detected via the $K^+K^-$ channel. In Ref.
\cite{chung97} we found that the inclusion of kaon medium effects
reduces the phi meson yield determined from the $K^+K^-$ analysis
by about a factor of two. Furthermore, the increase of phi meson
decay width broadens its mass distribution and results in a large
apparent width in the dilepton spectrum. The results in this
scenario are shown in Fig. \ref{eem}. As medium effects on vector
mesons are neglected, the background is the same as in Fig.
\ref{ee}. Because of the increase of phi meson decay width, the
dilepton mass spectrum from phi mesons becomes much broader than in
Fig. \ref{ee}, with a width of about 30-40 MeV. Consequently, the
height of the peak is substantially reduced, and the phi meson peak
is now below the background from $\pi\pi$ annihilation, leading to
a broad bump instead of a sharp peak around $m_\phi$ in the total
dilepton spectrum.

\begin{figure}
\begin{center}
\centerline{\epsfig{file=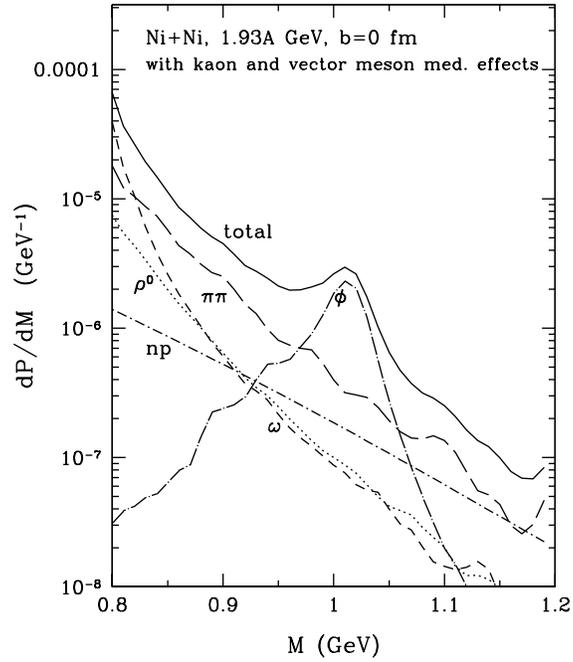,width=4in,height=4in}}
\caption{Same as Fig. \protect\ref{ee} for the scenario
including both kaon and vector meson medium effects.
\label{eemm}}
\end{center}
\end{figure}

The results in the case with both kaon and vector meson medium
effects are shown in Fig. \ref{eemm}. The most important background
around the phi meson is still from $\pi\pi$ annihilation, which is,
however, significantly smaller than in the case without vector
meson medium effects. This is mainly due to the dropping rho meson
mass that shifts the strength to masses below $m_\rho$. Although
the rho meson yield is enhanced due to its dropping mass, they
contribute mainly to dileptons with masses below $m_\rho$. On the
other hand, the contribution from direct omega meson decay is
enhanced in the dropping mass scenario, and it becomes comparable
to that from direct rho meson decay. This is due to the enhanced
production of omega meson, and the fact that most omega mesons
decay after the freeze-out, when their masses return to the free
ones. In this case the $\phi$ peak is about a factor of 3-4 above
the background. Furthermore, there appears a shoulder around the
invariant mass of 0.95 GeV, arising from the decay of phi mesons
inside the fireball. Unfortunately, the height of the shoulder is
about a factor of 2 below the background. In the total dilepton
spectrum, an outstanding and broad peak instead of a weak bump is
seen.

\begin{figure}
\begin{center}
\centerline{\epsfig{file=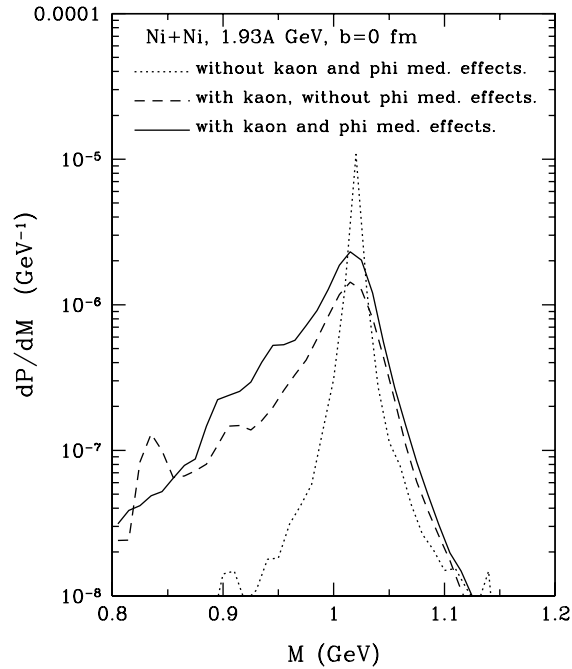,width=4in,height=4in}}
\caption{Dilepton invaraint mass spectra from phi meson decays in
central Ni+Ni collisions at 1.93A GeV.
\label{phiee}}
\end{center}
\end{figure}

In Fig. \ref{phiee} we summarize the dilepton spectra from phi
meson decay in the three scenarios discussed above. In the scenario
without medium effects, a very sharp phi peak is seen with a width
of about 4 MeV. Including kaon medium effects that increases the
phi meson decay width, the dilepton spectrum from phi meson decay
becomes much broader with a width of about 30-40 MeV, and its
height is substantially reduced. In the scenario with both
decreasing kaon and vector meson in-medium masses, the dilepton
spectrum is also quite broad with a width of about 30-40 MeV. Its
height increases, however, by about a factor of 2 with respect to
the second scenario, reflecting the fact that the phi meson yield
is increased due to a reduced mass. A shoulder, although not very
prominant, also develops around 0.95 GeV as a result of the decay
of phi mesons inside the fireball.

\begin{figure}
\begin{center}
\centerline{\epsfig{file=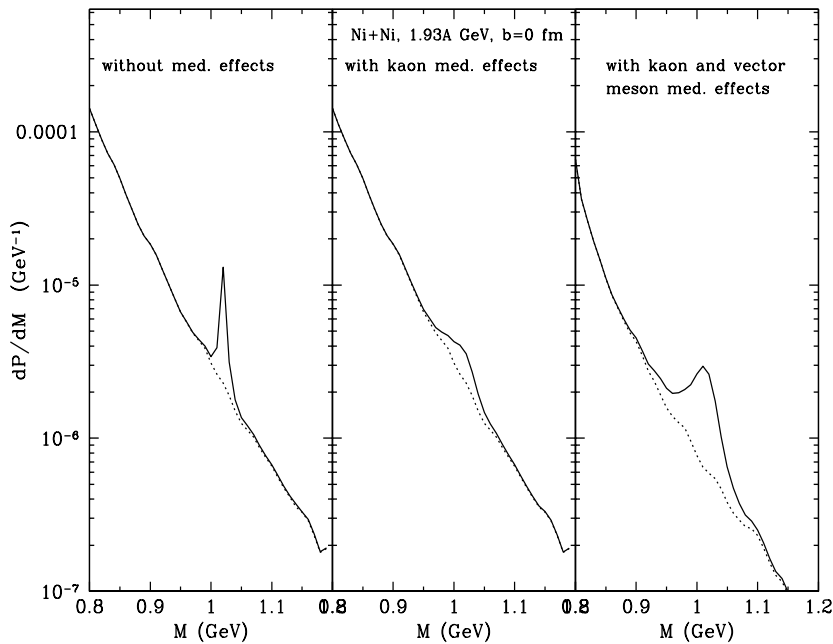,width=4in,height=5in,angle=270}}
\caption{Dilepton invaraint mass
spectra in central Ni+Ni collisions at 1.93A GeV. The dotted curves
give the total background.
\label{totee}}
\end{center}
\end{figure}

Our studies thus show that it will be quite difficult to isolate or
extract from the measured dilepton spectrum the contribution from
phi mesons. It is therefore useful to identify the characteristic
differences in the total dilepton spectra from the three scenarios.
This is shown in Fig. \ref{totee}. The dotted curves in the figure
show the background in the three scenarios, with those of the first
and second scenario being the same. With the HADES mass resolution
of about 1\%, we expect to see a sharp peak in the dilepton
spectrum which is about a factor of 5 above the background, if
there are no medium effects on both kaon and vector mesons. On the
other hand, a weak bump around $m_\phi$ would indicate that the phi
meson mass distribution becomes broader as a result of kaon medium
effects. Finally, if a broad and significant peak is seen around
$m_\phi$ together with a shoulder around 0.95 GeV, this could be a
good indication for the existence of both kaon and vector meson
medium effects.

\section{conclusion}

In conclusion, continuing our previous investigation on phi meson
production from heavy-ion collisions \cite{chung97}, we studied in
this paper the possibility of seeing the phi meson through its
dilepton spectrum. To make a quantitative prediction, we have
included contributions to dileptons with invariant mass around
$m_\phi$ from pion-pion annihilation, $pn$ bremsstrahlung, and the
direct decay of rho and omega mesons. The most important one is
found to come from pion-pion annihilation.

We have considered three scenarios for the kaon and vector meson
properties in nuclear medium. The dilepton spectra from phi meson
decays, as well as the total dilepton spectra, in the three
scenarios show quite different characteristics. We conclude that
with a mass resolution of about 1\% as to be achieved by the HADES
collaboration, it is possible to determine qualitatively whether
there are any medium effect on kaon and/or vector mesons. A
quantitative determination of the effective masses of kaons and
vector mesons from the dilepton spectrum is found to be difficult.
However, a combination of the information from the dilepton spectra
and other observables such as the phi meson yield from the analysis
of $K^+K^-$ channels as well as the kaon yields and spectra will
definitely provide good constraints on the in-medium properties of
kaons and vector mesons.

\begin{center}
\noindent {\bf Acknowledgement}
\end{center}

WSC and CMK were supported in part by the National Science
Foundation under Grant No. PHY-9509266 and The Welch Foundation
under Grant No. A-1358. GQL was supported in part by the Department
of Energy under Contract No. DE-FG02-88Er40388.

\end{document}